\documentclass[3p,times,procedia,number]{elsarticle}
\usepackage{ecrc}

\volume{00}

\firstpage{1}

\journalname{Procedia Engineering}

\runauth{Author name}

\jid{proeng}

\usepackage{amssymb}
\usepackage{graphicx,color,bm,amsmath}

 \biboptions{sort&compress}

\usepackage[figuresright]{rotating}

\usepackage[]{graphicx}
\graphicspath{{./figures/}{./figures/PDF/}{./figures/ipe/}}

\usepackage{amssymb}

\usepackage{amsthm}

\usepackage{lineno}

\usepackage{float}
\usepackage{multirow}
\usepackage{booktabs}
\usepackage{array}
\usepackage{xspace}   				
\usepackage{mathrsfs}
\usepackage{mdwlist}                 

\usepackage[english]{babel}
\usepackage[cyr]{aeguill}
\usepackage[ruled,vlined]{algorithm2e}

\usepackage[bordercolor=black,backgroundcolor=orange!60!white,linecolor=orange!60!white,figwidth=0.8\textwidth
]{todonotes}

\usepackage{bigints}
\usepackage{tensor}
\usepackage{multicol}
\usepackage{subfig}
\usepackage{tikz}
\usetikzlibrary{calc}
\usetikzlibrary{patterns}

\usepackage{listings}
\usepackage{xcolor} 

\usepackage[colorlinks=true,linkcolor=blue!20!black,urlcolor=black,citecolor=blue!20!black,hyperfootnotes=false]{hy
perref}

\usepackage[normalem]{ulem}

\usepackage{dsfont} 
\newcommand{\mesh}{\ensuremath{\mathcal{T}}}
\newcommand{\reals}{\ensuremath{\mathds{R}}}

\newcommand{\surface}{\ensuremath{\mathcal{S}}}
\newcommand{\inputtikzfile}[1]{%
\IfFileExists{tikz/#1.pdf}{\includegraphics[scale=1]{tikz/#1.pdf}}{\input{tikz/#1.tikz}}%
}

\definecolor{liquid}{rgb}{0.28,0.46,1}

\usetikzlibrary{external}
\begin{document}

\begin{frontmatter}



\dochead{26th International Meshing Roundtable}



\title{High quality mesh generation using cross and asterisk fields: Application on coastal domains.}


\address[a]{Universit\'e catholique de Louvain, MEMA, Avenue Georges Lemaitre 4, 1348 Louvain-la-Neuve, Belgium}

\author[a]{Christos~Georgiadis}
\author[a]{Pierre-Alexandre~Beaufort}
\author[a]{Jonathan~Lambrechts}
\author[a]{Jean-Fran{\c c}ois~Remacle}
\ead{jean-francois.remacle@uclouvain.be}

\begin{abstract}
This paper presents a method to generate high quality triangular or quadrilateral meshes that uses direction fields and a frontal point insertion strategy. Two types of direction fields are considered: asterisk fields and cross fields. With asterisk fields we generate high quality triangulations, while with cross fields we generate right-angled triangulations that are optimal for transformation to quadrilateral meshes. The input of our algorithm is an initial triangular mesh and a direction field calculated on it. The goal is to compute the vertices of the final mesh by an advancing front strategy along the direction field. We present an algorithm that enables to efficiently generate the points using solely information from the base mesh. A multi-threaded implementation of our algorithm is presented, allowing us to achieve significant speedup of the point generation. Regarding the quadrangulation process, we develop a quality criterion for right-angled triangles with respect to the local cross field and an optimization process based on it. Thus we are able to further improve the quality of the output quadrilaterals. The algorithm is demonstrated on the sphere and examples of high quality triangular and quadrilateral meshes of coastal domains are presented.
\end{abstract}


\end{frontmatter}





\section{Introduction}

This work is motivated by the increasing use of unstructured meshes in ocean modelling.
Unstructured meshes offer certain advantages in the context of geophysical simulations, such as conforming to the complex geometries of coastal domains and supporting variable mesh sizes \cite{Candy2017,Engwirda2015}. One of the main difficulty in dealing with geographical data is the over-sampled nature of coastline representations. In \cite{remacle2016fast} we propose an
algorithm that automatically unrefines coastline data and we provide a fast and robust mesh generation procedure that is able to generate meshes of the earth system (ocean and continent).
However, this method only generates triangular meshes and is specific to the sphere. For geophysical flow simulations quadrilateral meshes are desirable, since they contain twice less elements and can be aligned with the actual flow characteristics.
 
In this paper we introduce a general method to generate high quality quadrilateral and triangular meshes of surfaces. The input of our algorithm is an initial mesh $\mesh_0$, in particular a mesh on the sphere generated with the method proposed in \cite{remacle2016fast}. The output is either a high quality triangular mesh or a right-angled triangular mesh that is transformed into an quadrilateral mesh. Both the description of the geometry and the Delaunay kernel are borrowed from  \cite{remacle2016fast}. There, the point insertion strategy was based on a basic 
edge saturation Delaunay refinement procedure. The main contribution of this work is a new strategy to generate the vertices of the final mesh in an optimal way. By optimal we mean that the generated Delaunay triangles can be recombined in high quality quadrilaterals or that they are as equilateral as possible. 

The method has similarities with \cite{baker1988nonobtuse} or 
\cite{bern1994provably} where authors use uniform grids or balanced
quadtrees to locate the internal vertices of the mesh. The drawback of
such methods is that quadtrees are axis oriented and meshes may be of
low quality close to boundaries. 
The method that is proposed here overcomes these challenges. 

All vertices are inserted using a frontal approach before computing their Delaunay triangulation. To steer the insertion of points we utilize direction fields. For quadrilateral meshing, the direction field is a cross field \cite{bommes2009mixed,kowalski2013pde,remacle2013frontal}, while for triangular meshing the direction field is an asterisk field \cite{PAIMR26} (Section \ref{sec::dirfields}). This direction field is computed on the base mesh $\mesh_0$  using a Ginzburg-Landau formulation \cite{PAIMR26}. To compute the coordinates of the vertices we are using an efficient intersection strategy that does not rely on any global geometrical structures, e.g. octrees (Section \ref{sec::intersection}). Similarly, filtering of close points is done locally on the fly (Setion \ref{sec::filtering}). The whole procedure can be parallelized with a multi-threaded strategy proposed in Section \ref{sec::parallel}. Finally, a novel way to measure the quality of right-angled triangles is proposed and utilized to optimize output meshes (Section \ref{sec::recombination}) for combination into quads with the blossom-quad algorithm \cite{remacle2012blossom}.

The method is demonstrated here on spherical geometries. We present results for triangular and quadrilateral meshes on the world ocean (Section \ref{sec::results}). 
Nevertheless, the algorithm is designed and implemented as a tool that
can handle triangulations of any orientable surface $\surface$ embedded in $\reals^3$. 

The method presented has been released as a self
consistent open source code that can be used as a stand-alone program
or that can be plugged in other software's such as Gmsh \cite{geuzaine2009gmsh} or
QGIS \cite{qgis2011quantum}. 

\section{Frontal point insertion}
To begin, we consider a base mesh $\mesh_0$ of an orientable manifold surface.
The objective is to spawn points inside the domain in preferred directions. 
We essentially want to obtain a set of points $\ensuremath{P}=
\{\mathbf{p}_1,\ldots,\mathbf{p}_n\}$ that will either 
(i) lead to a triangulation $\mesh_q$ that is well suited for combining triangles
into quadrilaterals or (ii) lead to a triangulation $\mesh_t$ that contains triangles that
are close to equilateral. The idea is to use direction fields of two different kinds, depending on whether one wants to
generate $\mesh_q$ or $\mesh_t$.

\subsection{Direction fields}\label{sec::dirfields}

A \emph{cross field} ${\bf c}$ is a field defined on a surface $\surface$
with values in the quotient space $S^1/Q$, 
where $S^1$ is the circle group  and $Q$ is the group of quadrilateral symmetry. 
It associates to each point of a surface $\surface$ to be meshed
a cross made of four unit vectors orthogonal to one another in the
tangent plane of  the surface. In the context of quadrilateral mesh generation, a
cross field represents at each point of the domain the preferred 
orientations of a quadrilateral mesh. 

It is possible to build direction fields for building $\mesh_t$ by
defining asterisk fields. An \emph{asterisk field} ${\bf a}$ is also 
a field defined on a surface $\surface$ but with values in the quotient space $S^1/H$, 
where $H$ is the symmetry group of a regular hexagon. Pictorially, it
associates to each point of the surface $\surface$
an asterisk made of six unit vectors separated by $60$ degrees. Such a
field will be used to align triangular meshes to pre-defined
directions in order to build triangles that are close to equilateral.

Direction fields should be as smooth as possible and should be
aligned with the boundary of the domain. One can show
that smooth fields only exist for surfaces with a
Poincar\'e characteristic equal to $0$. For any other topology,
critical points will necessarily occur. More details on direction fields
and their computation can be found in \cite{PAIMR26}.

\begin{figure}
\begin{center}
\includegraphics[width=6cm]{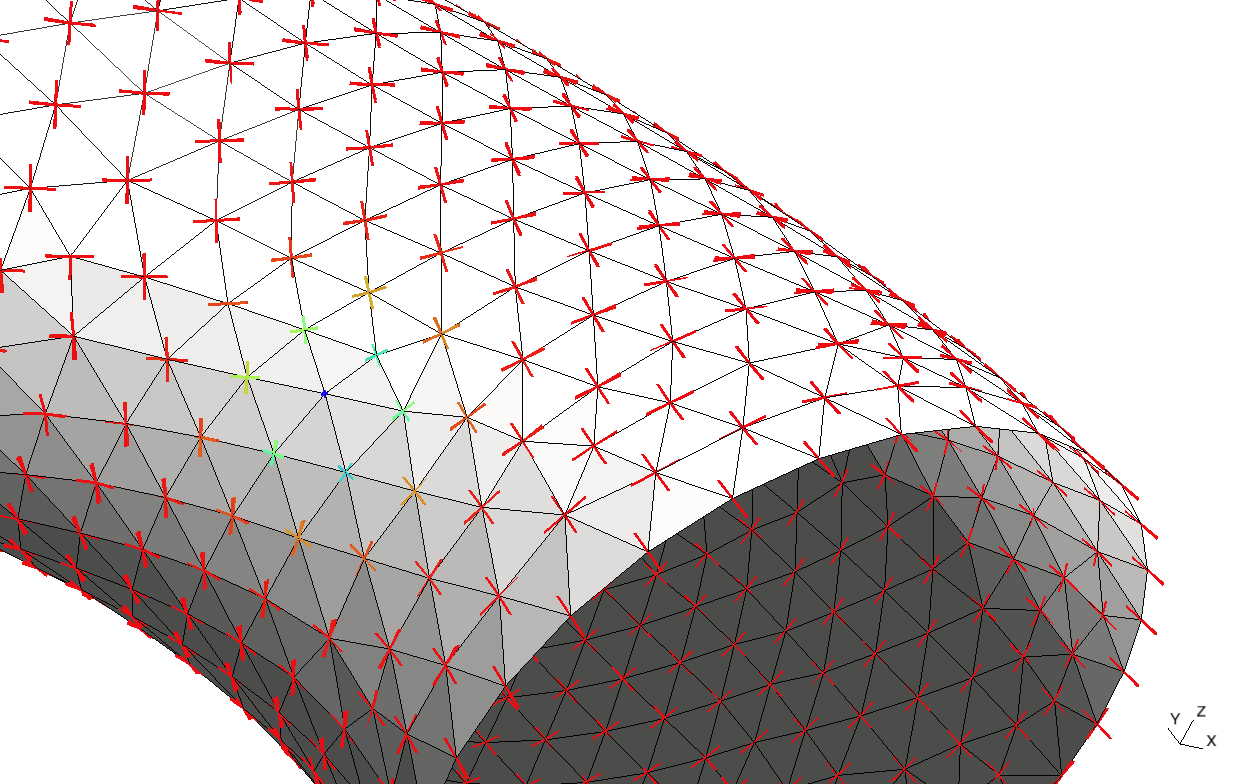}
\includegraphics[width=6cm]{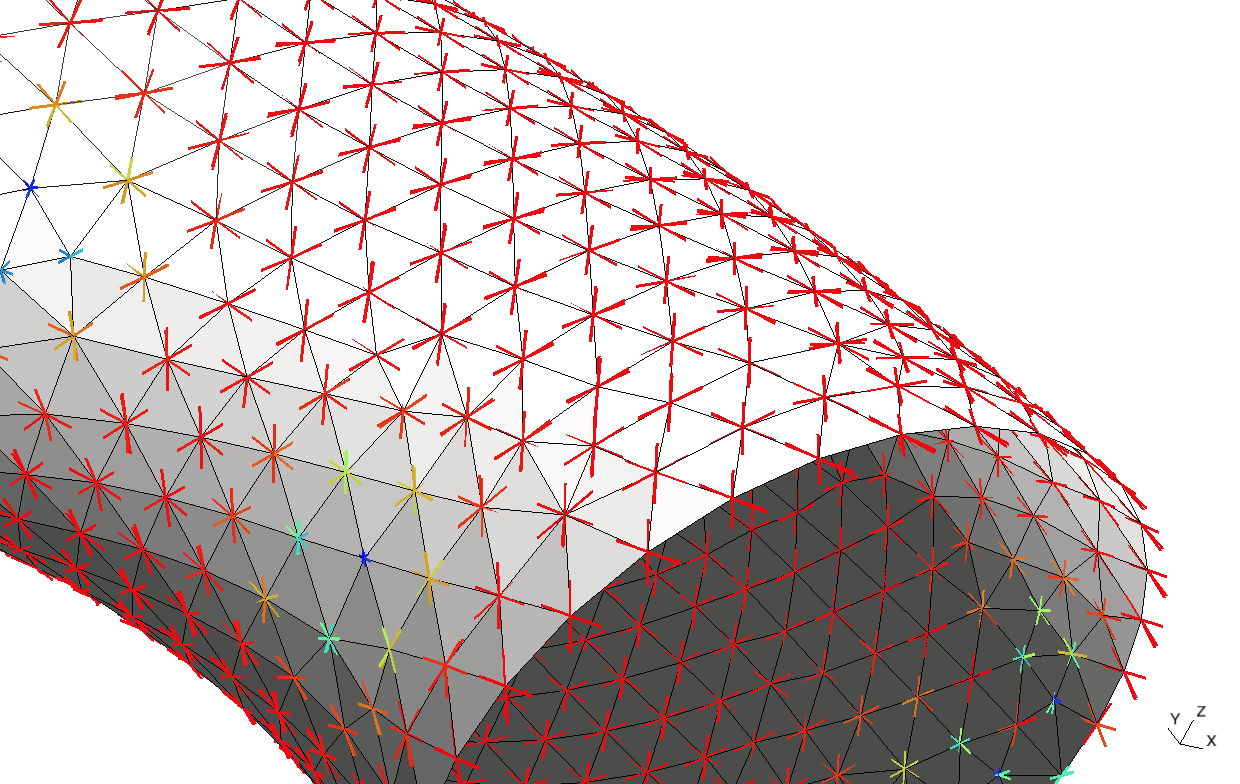}
\end{center}
\caption{Cross field (left) and Asterisk field (right)\label{fig:cross}}
\end{figure}

A cross field and an asterisk field are presented in Figure \ref{fig:cross}.

\subsection{Generating points across direction fields}

Assume a direction field ${\bf f}$ defined everywhere on
$\surface$. Field $\bf f$ is either a cross field ($N_d = 4$ directions)
or an asterisk field ($N_d = 6$ directions). 
A priority queue is initially filled with all the boundary points of the base mesh$\mesh_0$. 
The point ${\bf p}_i$ at the top of the queue then tries to insert $N_d$
points ${\bf p}_{ij}$  in the $j=1,\dots,N_d$ directions defined 
by ${\bf f}({\bf p}_i)$ and at a distance $h({\bf p}_i)$
where $h$ is the mesh size field. In order to have points inserted ``by
layers'', the priority queue that is chosen is a first-in, first-out queue.
Our experience shows that ordering the 1D points allows smooth
propagation of the fronts inside the domain. We have tested  two kind
of ordering: (i) order the points geometrically by sorting them along
a space filling curve (Hilbert curve) and (ii) order the points
topologically by walking along the 1D boundary. 

Each point $\mathbf{p}_{i}$ thus tries to spawns $\mathbf{p}_{ij},$ $j=1,\ldots,N_d$
neighbour points on $\mathcal{T}_0$. Yet, there is no guarantee that
point $\mathbf{p}_{ij}$ is not too close to another point of
the queue. Points $\mathbf{p}_{ij}$ are hence filtered. An exclusion zone of size
$\alpha h(\mathbf{p}_i)$, $\alpha < 1$  is created around every vertex $\mathbf{p}_i$
of the queue in such a way that no new point can be inserted in the
queue if it lies inside this zone. Finally, accepted points are added to the end of the queue
as well  as in  $\ensuremath{P}$. The procedure terminates when the 
queue is empty. Algorithm \ref{sec:fpi} describes the procedure. In the two following subsections we will focus with detail on the two main operations, i.e. the point insertion and the point filtering.

\label{sec:fpi}
\begin{algorithm}
     \SetKwData{Left}{left}\SetKwData{This}{this}\SetKwData{Up}{up}
     \SetKwFunction{Union}{Union}\SetKwFunction{FindCompress}{FindCompress}
     \SetKwInOut{Input}{input}\SetKwInOut{Output}{output}   
     \SetKw{KwBy}{by}
   
      \Input{Initial triangulation $\mesh_0$, mesh size field function $h({\bf x})$ and a direction field calculated on $\mesh_0$}
      \Output{Array of points in preferred directions $\ensuremath{P}$ to be triangulated.}
      \BlankLine
      Place boundary points in a queue\;
      Create data structure to store in which triangle of $\mesh_0$ the generated points lie\;
      \BlankLine
      \While{queue is not empty} {
         take the first point $\mathbf p_{i}$ at the top of the queue\;
         pop this point out of the queue\;
         interpolate the direction vectors for this point\; 
		 \For{$N_d$ directions}{
    		Insert point $\mathbf{p}_{ij}$ by intersecting the base triangulation (Section \ref{sec::intersection})\;
    		create set of triangles $\mathcal{C}$ in exclusion zone around $\mathbf{p}_{ij}$ to filter (Section \ref{sec::filtering})\;
    		\For{$\mathbf p_{k} \in \mathcal{C}$}{
    		\If{$\|\mathbf{p}_{ij} - \mathbf p_{k}\| > \alpha h(\mathbf{p}_{ij})$ }{
    			add $\mathbf{p}_{ij}$ in $\ensuremath{P}$\;
    			push $\mathbf{p}_{ij}$ in the back of the queue\;
    	    }
    	 	\Else{
    	    	delete $\mathbf{p}_{ij}$\;
    	 	}
    	 	}
         }        

     }
   
\caption{Frontal point insertion algorithm.}
\label{sec:fpi}
\end{algorithm}

\subsection{Intersection with triangulation}\label{sec::intersection}

\begin{figure}
\begin{center}
\includegraphics[width=0.5\textwidth]{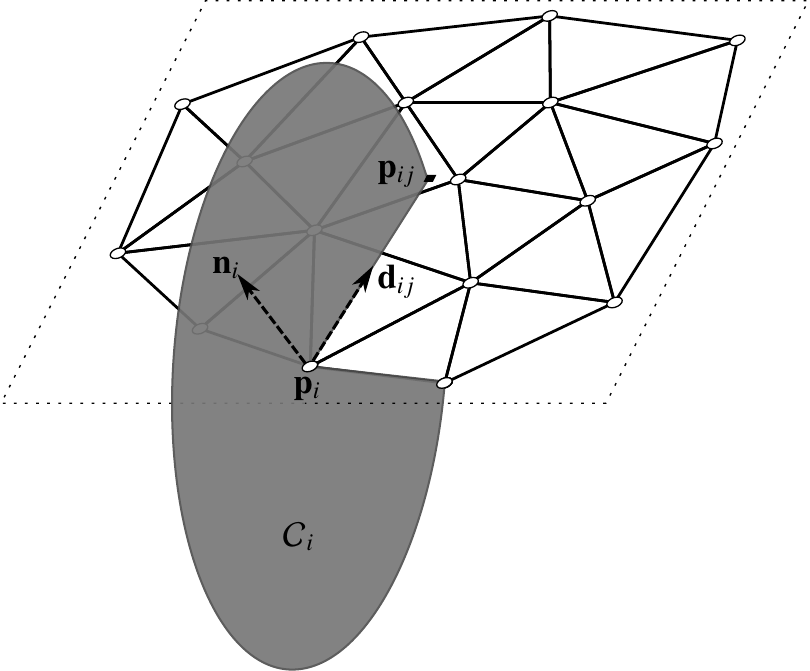};
\caption{Computation of point $\mathbf d_{ij}$.}
\label{fig::intersection}
\end{center}
\end{figure}

Assume a point $\mathbf{p}_{i}$ that lies on one of the triangles of $\mesh_0$, 
a direction $\mathbf{d}_{ij}$ i.e. a unit vector tangent to the
surface and the mesh size $h(\mathbf{p}_{i})$ at that point. The aim
is obviously to create an edge of size $h$. So, the new point
$\mathbf{p}_{ij}$ will be located at the intersection of the
triangulated surface $\mesh_0$ and a circle $\mathcal{C}_i$ of center $\mathbf{p}_i$ and radius $h(\mathbf{p}_i)$. $\mathcal{C}_i$ lies on the plane $\mathcal{P}_i$ that is formed by the direction vector $\mathbf{d}_{ij}$ and the normal to the triangulation at our origin point, $\mathbf{n}_i$ (Figure \ref{fig::intersection}).
To compute $\mathbf{p}_{ij}$ our goal is to find the intersection point of circle $\mathcal{C}_i$  with the triangulation $\mesh_0$ (Figure \ref{fig::intersection}). 
We start from the triangle of the base mesh $\mesh_0{}_{i}$ on which $\mathbf p_{i}$ lies.
First, we compute the intersection line of the plane $\mathcal{P}_i$ and the plane of the triangle $\mathcal{P}_{\mesh_0{}_{i}}$.
Then, we find the intersection points of this line with the circle $\mathcal{C}_i$ and choose the one that lies in direction $\mathbf{d}_{ij}$.
Finally, the barycentric coordinates $\lambda_0,\lambda_1,\lambda_2$ of this point with respect to the current triangle are calculated.
In this way we determine whether the intersection point lies on the triangle, and thus a successful intersection with this triangle. 

\begin{figure}
\centering
\begin{center}
\begin{minipage}{.4\textwidth}
\centering
\includegraphics[width=0.8\textwidth]{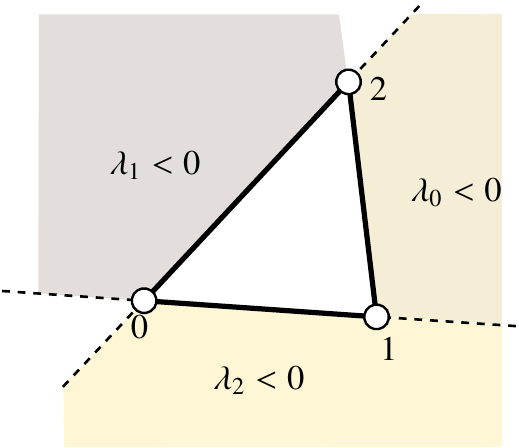};
\end{minipage}
\begin{minipage}{.4\textwidth}
\centering
\includegraphics[width=0.8\textwidth]{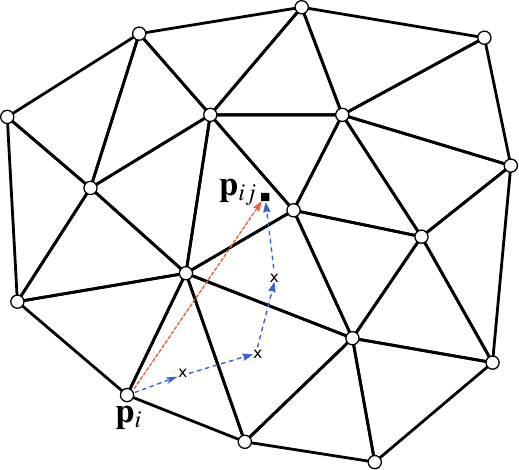};
\end{minipage}
\caption{Indication of the position of intersection point according to barycentric coordinates (left) and computation of point $\mathbf p_{ij}$ by walking in the triangulation in specific direction (right) }
\label{fig::walking}
\end{center}
\end{figure}

In the case where the current triangle is not intersected, we move forward to another triangle. 
Since we have already computed the barycentric coordinates with respect to the current triangle $\mesh_0{}_{i}$, we know where on the plane $\mathcal{P}_{\mesh_0{}_{i}}$ the intersection point lies  (Figure \ref{fig::walking}, left). Therefore, we have an indication on which neighbour triangle we should search after. This procedure continues until a valid intersection point is retrieved. 

Essentially, we perform a walk in the triangulation \citep{Devillers2002} in the desired direction, until we obtain the intersection point. Our experience shows that this method is the most  efficient one amongst all possible solutions for this problem. For same mesh size field chosen for the initial and the output mesh, intersection points are found after a little less than 2 triangle visits on average. 

\subsection{Filtering}\label{sec::filtering}

As it was previously explained, each point generates $\mathbf{p}_{ij}$ points for $j=1,\ldots,N_d$ directions. 
We have to ensure that new points are not too close to already generated points.
Therefore, after each point $\mathbf{p}_{ij}$ is generated, a filtering procedure should follow.
We prefer not to utilize a space search structure such as an RTree \cite{beckmann1990r}, since that would make parallelization more challenging.

\begin{figure}
\begin{center}
\includegraphics[width=0.4\textwidth]{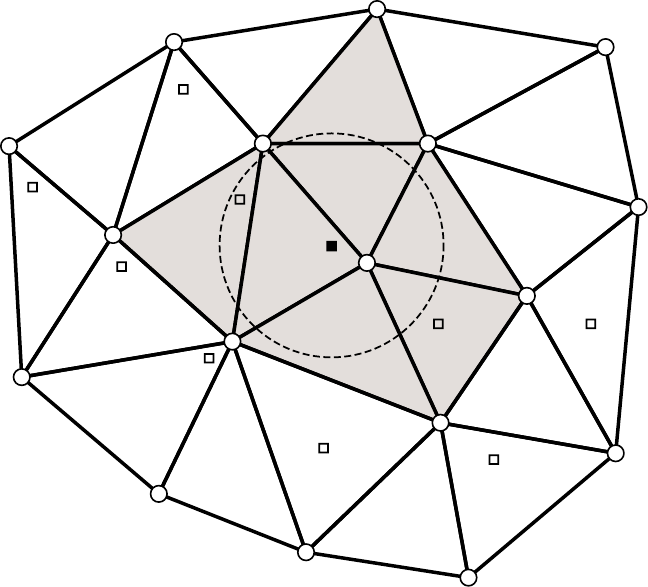};
\captionsetup{justification=centering}
\caption{Filtering procedure for new points.$---$: exclusion zone, $\mathcal{C}$: triangles intersecting the exclusion zone in light grey,\\ $\blacksquare$: candidate point, $\square$: previously generated points stored on $\mesh_0$}
\label{fig::filtering}
\end{center}
\end{figure}

Instead, we proceed with a more straightforward strategy.
During the point insertion process, we store on which triangle of the base mesh $\mesh_0$ each inserted point lies.  
For every candidate point $\mathbf{p}_{ij}$, we take the set of triangles $\mathcal{C}$ around our new point that intersect an exclusion zone (Figure \ref{fig::filtering}).
The exclusion zone is defined as a circle of radius $h(\mathbf{p}_{ij})$ around the candidate point.
Therefore we are able to directly obtain the set of points $\ensuremath{P}_{f} = \{\mathbf p_{k},k=1,\ldots,n_{f}\}$ in the vicinity of $\mathbf{p}_{ij}$. 
Since our objective is to create right-angled triangles, i.e. equilateral triangles in the $\mathcal{L}_{\infty}$ norm, we compute the distance between the candidate point and its surrounding ones as
$
\|\mathbf{p}_{ij} - \mathbf p_{k}\|_{\infty} = \text{max}
\{ |x_{ij}-x_k|,|y_{ij}-y_k|,|z_{ij}-z_k| \}
$
The point is accepted for insertion if condition 
$
\|\mathbf{p}_{ij} - \mathbf p_{k}\|_{\infty} > \alpha \cdot h(\mathbf{p}_{ij})
$
holds for all $\mathbf{p}_k \in \ensuremath{P}_{f}$.\\

\subsection{Parallel implementation}\label{sec::parallel}

A multi-threaded Delaunay kernel, as well as a multi-threaded Delaunay refinement process based on it, have already been presented \cite{remacle2016fast,remacle2015} and can significantly reduce the time cost of the mesh generation process. The frontal point generation algorithm that was presented could add up to being the most significant term in the total timings of the mesh generation workflow. Therefore, and in order to combine with the parallel Delaunay triangulation process, the generation of points in a multi-threaded fashion would be highly desirable.

Due to the locality of point insertion and filtering, the whole process can be parallelized with the following straightforward strategy. As before, the boundary points are sorted along the boundary elements of $\mesh_0$. This set of initial points is then split in $M$ queues, where $M$ is the number of threads. An OMP parallel region is initiated, where each thread $m=1,\ldots,M$ handles its own subset of points independently. The threads have shared memory access to the final set of points $\ensuremath{P}$ and the structure that stores the points for each triangle of $\mesh_0$. Each point $\mathbf{p}_{i}^m$ generates $\mathbf{p}_{ij}^m$ , $j=1,\ldots,N_d$ ($N_d=4$ or $N_d=6$) new points around it. These points are filtered out with respect to the already inserted points. Accepted points in the local filtering process are flagged, but not yet inserted in the respective data structures.
   
In order to ensure that different threads do not insert points close to each other and that do not access the global data structures at the same time, an OMP critical region follows. There, the candidate point $\mathbf{p}_{ij}^m$ of the current thread is checked against the already accepted points from the other threads at this point. If no other point is near, the point gets accepted and the procedure goes on. The process terminates when all $M$ queues are empty.

Initial numerical tests show that we can have a significant speedup with the aforementioned strategy. Currently, the algorithm can generate 1 million points in around 10 seconds in 1 thread, while for 4 threads the time decreases to around 3.4 seconds.
 
\section{Recombination to quadrilaterals}\label{sec::recombination}

\subsection{Right-angled quality criterion}\label{sec::quality}

\begin{figure}
\centering
\begin{minipage}{.30\textwidth}
\begin{flushleft}
\includegraphics[width=0.75\textwidth]{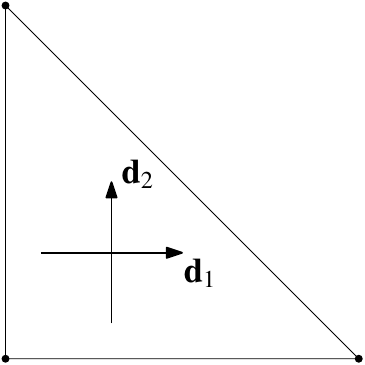};
\end{flushleft}
\end{minipage}
\begin{minipage}{.30\textwidth}
\centering
\includegraphics[width=0.75\textwidth]{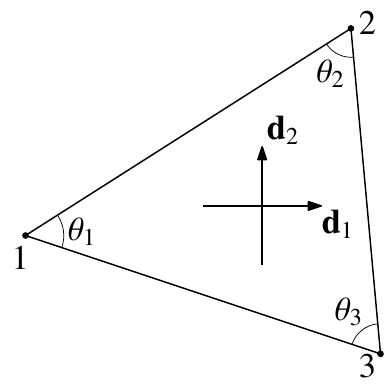};
\end{minipage}
\begin{minipage}{.30\textwidth}
\begin{flushright}
\includegraphics[width=0.7\textwidth]{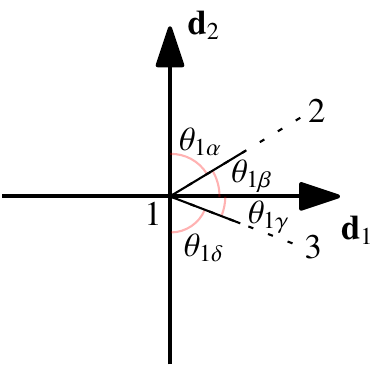};
\end{flushright}
\end{minipage}
\caption{Optimal right-angled triangle aligned with the cross frame (left), triangle to be evaluated (middle), angles between edges of vertex 1 with the two main directions (right)}
\label{fig::quality}
\end{figure}

Points along a cross field will generate right-angled triangles that are suitable for merging into quadrilaterals. Our concern now is to further improve the quality of these triangles before applying a recombination procedure. 
We consider that the optimal triangles would be the right-angled ones in respect to the local cross frame (Fig. \ref{fig::quality}, left), since they can be combined in quadrilaterals of optimal quality.
To this end, we are investigating the development of a quality criterion for right-angled triangles, based on how 'close' the triangle is to the optimal one.\\

For each vertex of the triangle ($i=1,2,3$) we calculate three corresponding local qualities, represented from the following dimensionless quantities:

\begin{itemize}
\item[(i)] a quantity to evaluate how close each angle of the triangle is to $90^{\circ}$:
$$ 
q_{a}^{i}=1-\dfrac{|\frac{\pi}{2}-\theta_i|}{\frac{\pi}{2}} 
$$
 
\item[(ii)] a quantity to evaluate the nearest edge of each vertex to one direction of our the local cross field. For this, we calculate for the two edges of each vertex their angle with the two main directions $\mathbf{d}_{1}$ and $\mathbf{d}_{2}$, thus $j=4$ angles (Fig. \ref{fig::quality}, right), and we choose the smallest one:
$$
q_{b}^{i}= \text{max} \Big\lbrace 
	\vert \cos(2 \theta_{ij}) \vert \ : \ j=\alpha,\beta,\gamma,\delta \Big\rbrace 
$$

\item[(iii)] finally, the ratio of the two edges of each vertex:
$$
q_{c}^{i}= 1 - \dfrac{\vert e_{ik} - e_{ik} \vert}{\text{max}(e_{ik},e_{ik})} \ , \
					 k=\lbrace 1,2,3 \ : \ k\neq i \rbrace
$$

the optimal would be that the edges have equal length, i.e. it is a right-angled triangle, equilateral in the $\mathcal{L}_{\infty}$ norm. If the objective is to obtain anisotropic quadrangles through the merging of anisotropic right-angled triangles, this quantity can be easily modified to take into account the specified size ratio.
\end{itemize}

The final right-angled quality of the triangle is the maximum amongst the product of the three local qualities of its vertices: $
q_t = \text{max} \Big\lbrace (q_{a} \cdot q_{b} \cdot q_{c})^{i}  \  : \  i=1,2,3 \Big\rbrace
$. 

\subsection{Optimize for recombination to quadrilaterals}\label{sec::optimization}

\begin{figure}[H]
\begin{center}
\includegraphics[width=0.4\textwidth]{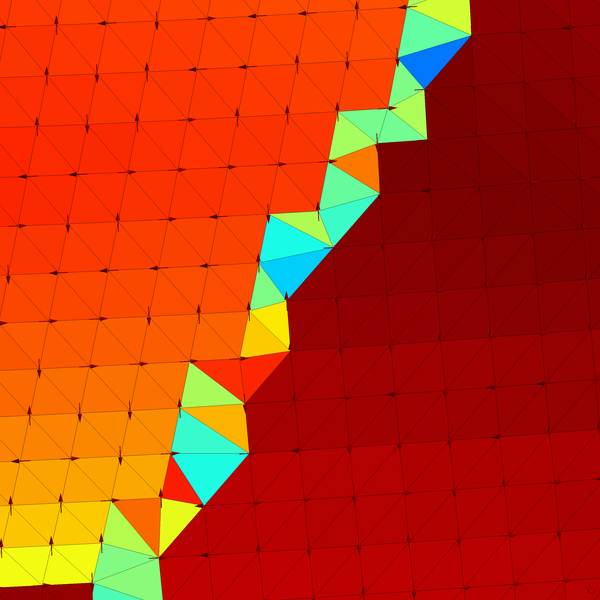}
\includegraphics[width=0.4\textwidth]{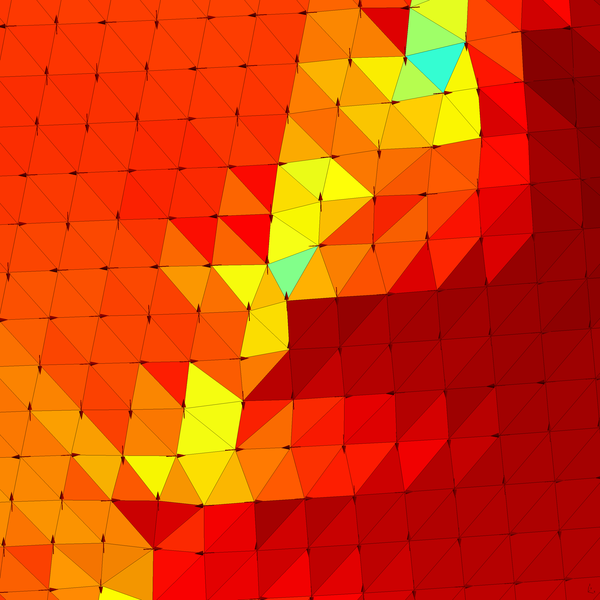}
\includegraphics[width=0.6\textwidth]{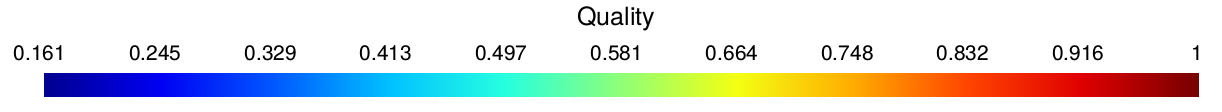}
\caption{Quality distribution on an interface where fronts of insertion collide (one originating from the west and one from the east): before (left) and after (right) optimization. Low quality triangles are transformed and a smooth transition with pairs of right-angled triangles is obtained. \label{fig::qualoptim}}
\end{center}
\end{figure}

We have now a measure to evaluate the quality of right-angled triangles. We expect that in certain regions the generated points will not create triangles optimal for recombination.
Such regions are located for example in parts of the domain where there is transition to different mesh sizes or where fronts of insertion collide (Fig. \ref{fig::qualoptim}, left). Herein we follow a simple procedure in order to optimize these regions, based on the quality measure presented in section \ref{sec::quality}.

For each interior point of the domain we take its corresponding cavity of surrounding triangles. The minimum quality amongst the cavity's triangles is set to be the objective function to be maximized. A local maximum is then searched in the line connecting the original position of the point and the centroid of the cavity in the infinity norm, and the point is relocated accordingly. The aforementioned optimization process is done for all the interior cavities of the mesh until points are not relocated further. The improvement of the right-angled quality of triangles in an insertion front collision is depicted on Figure \ref{fig::qualoptim}. 

\section{Results}\label{sec::results}

\subsection{Triangulations}\label{sec::restri}

\begin{figure}[H]
  \begin{center}
  \begin{tabular}{cc}
    \setlength{\fboxsep}{0pt} \fbox{\includegraphics[width=7cm]{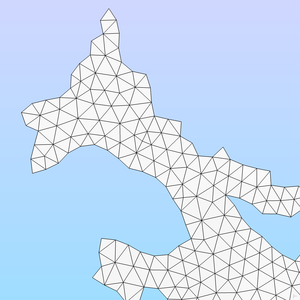}}& 
    \setlength{\fboxsep}{0pt} \fbox{\includegraphics[width=7cm]{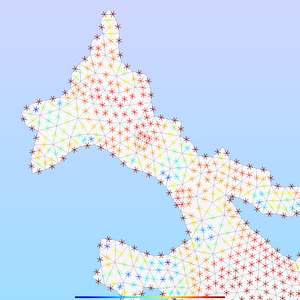}}\\
    \footnotesize{(a) Base mesh (detail of Baltic sea) } &
    \footnotesize{(b) Asterisk field calculated on base mesh} \\ [0.3cm]
    
    \setlength{\fboxsep}{0pt} \fbox{\includegraphics[width=7cm]{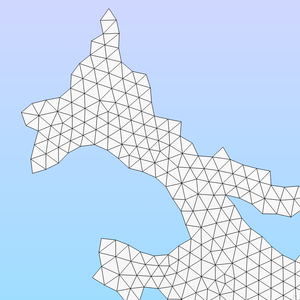}}&
    \setlength{\fboxsep}{0pt} \fbox{\includegraphics[width=7cm]{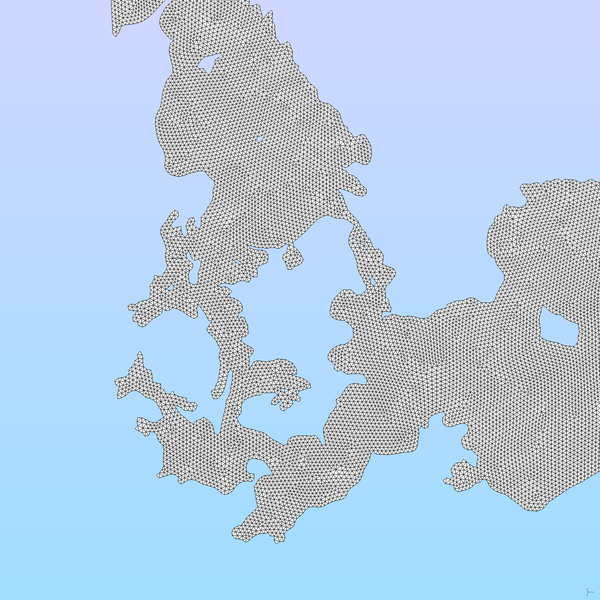}}\\
    \footnotesize{(c) Final mesh} &
    \footnotesize{Wider view of final mesh} 
  \end{tabular}
  \caption{Generating a high quality triangular mesh by using an asterisk field. \label{fig::triangular}}
  \end{center}
\end{figure}

\begin{figure}
  \begin{center}
  \begin{tabular}{cc}
    \setlength{\fboxsep}{0pt} \fbox{\includegraphics[width=7cm]{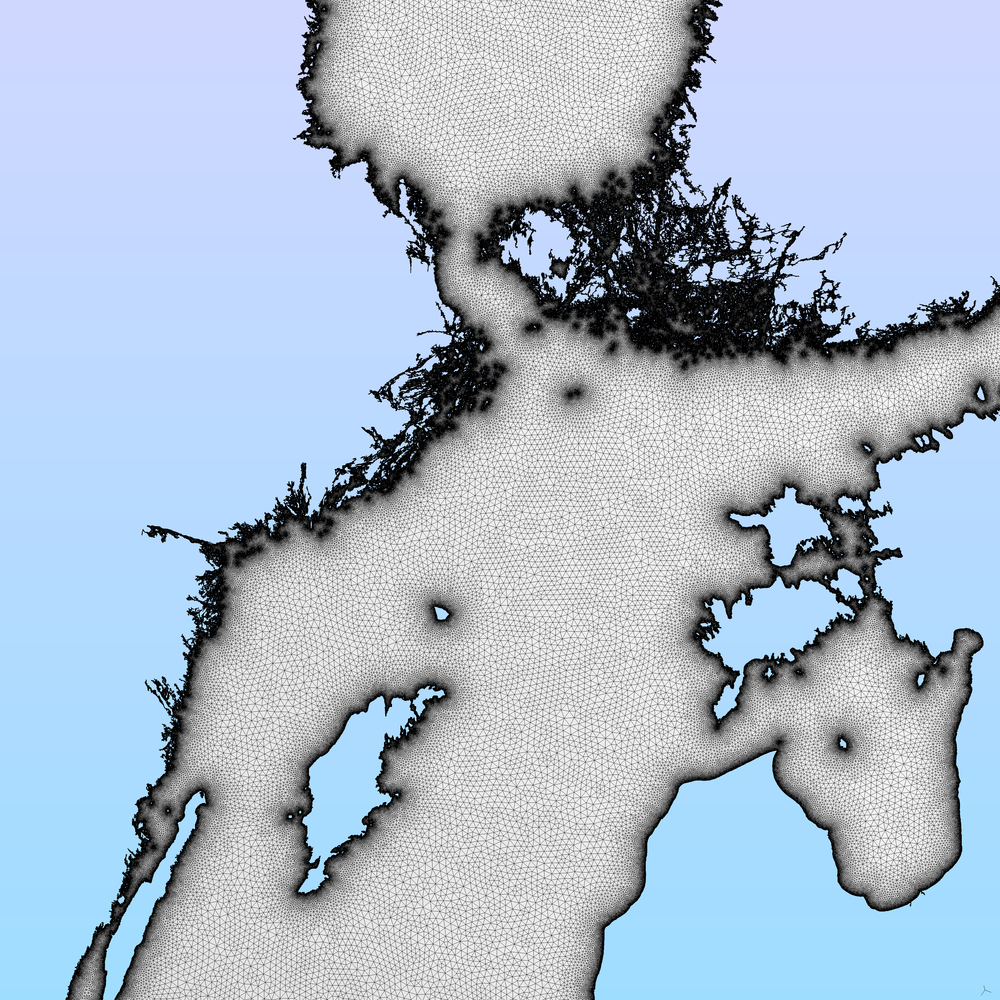}}&         
    \setlength{\fboxsep}{0pt} \fbox{\includegraphics[width=7cm]{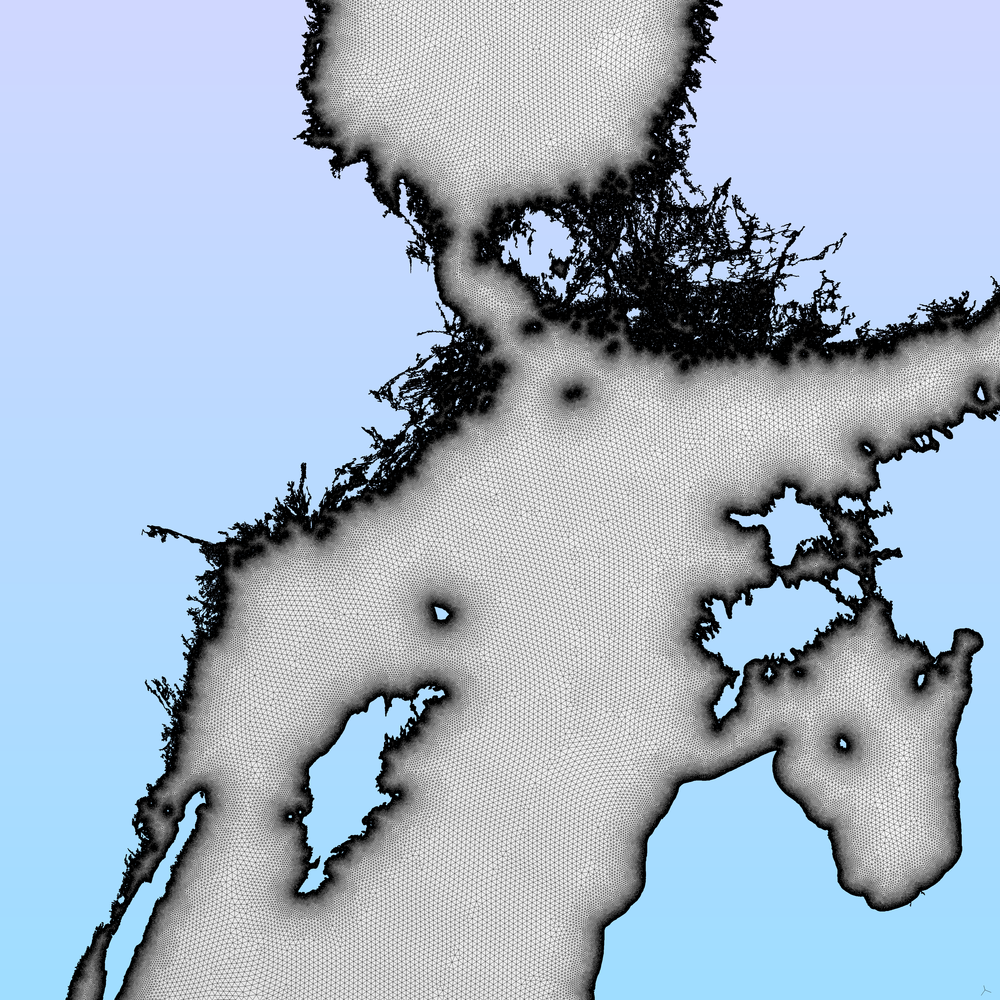}}\\
    
    \setlength{\fboxsep}{0pt} \fbox{\includegraphics[width=7cm]{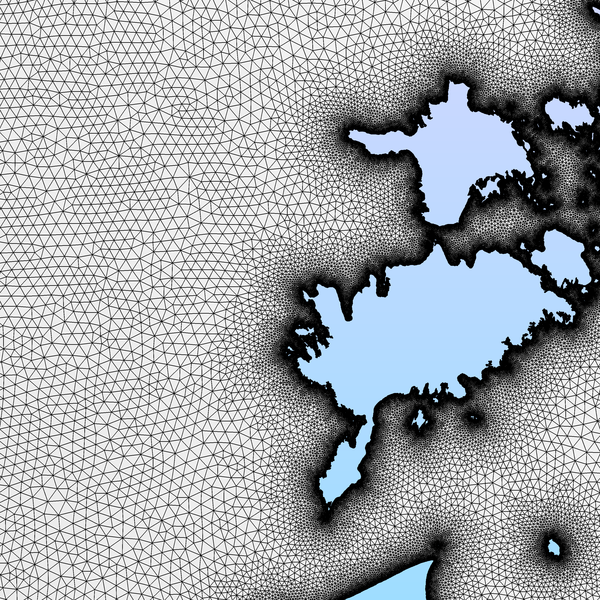}}&     
    \setlength{\fboxsep}{0pt} \fbox{\includegraphics[width=7cm]{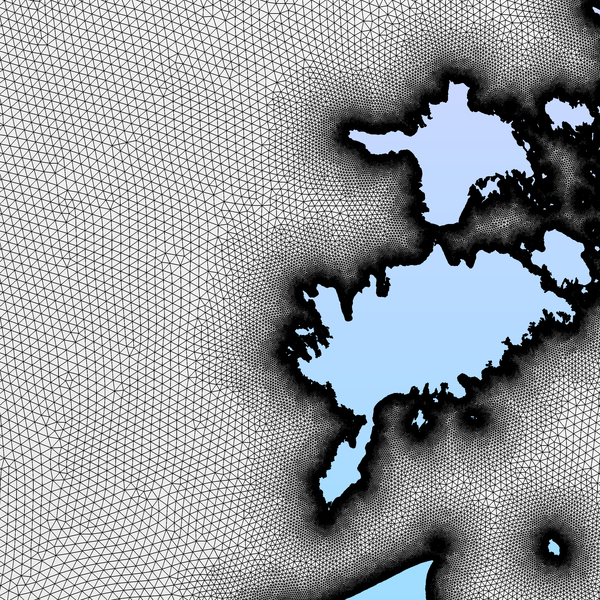}}\\
        
    \setlength{\fboxsep}{0pt} \fbox{\includegraphics[width=7cm]{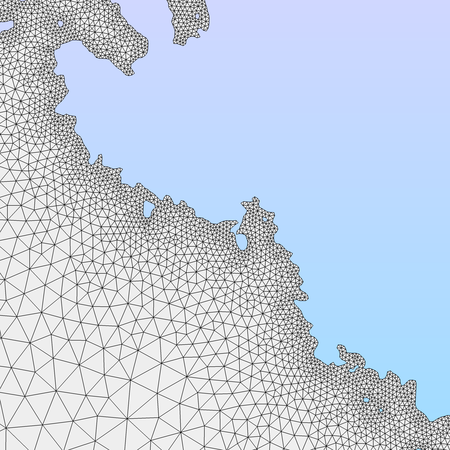}}&            
    \setlength{\fboxsep}{0pt} \fbox{\includegraphics[width=7cm]{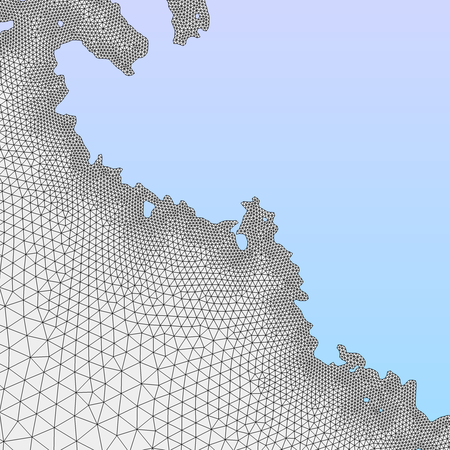}}\\    

  \end{tabular}
  \caption{Images of base (left) and output (right) triangulation on the Baltic sea. Average        radius-ratio quality (defined as $\gamma = \dfrac{2r_i}{r_c}$, where $r_i$ and $r_c$ are the radii of the inscribed and circumscribed circles of the triangle respectively) is $\overline{\gamma} = 0.947$ ($\gamma_{min}=0.104$) for the initial mesh. The output mesh improves it to $\overline{\gamma} = 0.981$($\gamma_{min}=0.0.3231$). \label{fig::tribaltic}}
  \end{center}
\end{figure}

We present here results for mesh generation by using asterisk fields. Figure \ref{fig::triangular} illustrates the process for generating a triangular mesh in such a way.  In Figure \ref{fig::tribaltic} we present a comparison between the base and the output mesh on Baltic sea, choosing a high resolution mesh size field: from $h_{min}=150m$ on the coast to $h=3km$ away from it. The final triangular mesh exhibits a smoother distribution of equilateral triangles through the domain.

\subsection{Quadrangulations}\label{sec::resquad}

\begin{figure}
  \begin{center}
  \begin{tabular}{cc}
    \setlength{\fboxsep}{0pt} \fbox{\includegraphics[width=8cm]{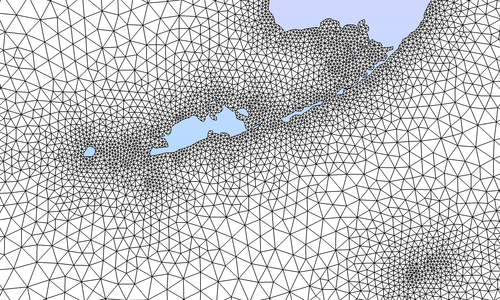}}& 
    \setlength{\fboxsep}{0pt} \fbox{\includegraphics[width=8cm]{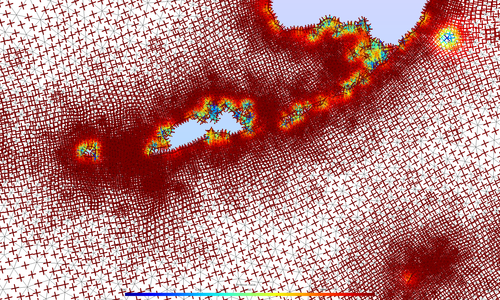}}\\
    \footnotesize{(a) base mesh (white is water)} &
    \footnotesize{(b) cross field on base mesh} \\ [0.3cm]
    
    \setlength{\fboxsep}{0pt} \fbox{\includegraphics[width=8cm]{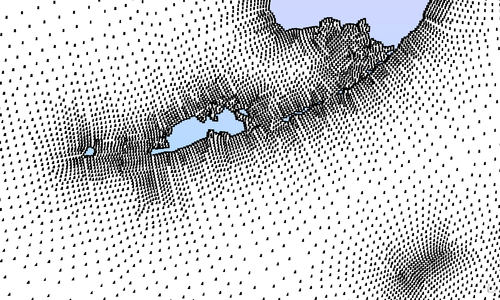}}&
    \setlength{\fboxsep}{0pt} \fbox{\includegraphics[width=8cm]{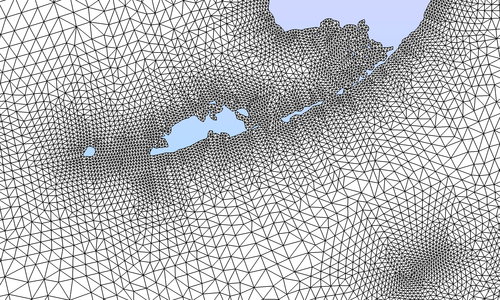}}\\
    \footnotesize{(c) generation of points in the domain} &
    \footnotesize{(d) right-angled triangulation} \\ [0.3cm]
    
    \setlength{\fboxsep}{0pt} \fbox{\includegraphics[width=8cm]{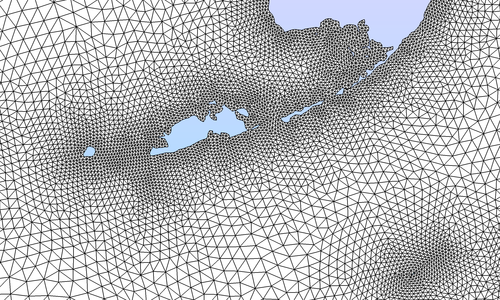}}&
    \setlength{\fboxsep}{0pt} \fbox{\includegraphics[width=8cm]{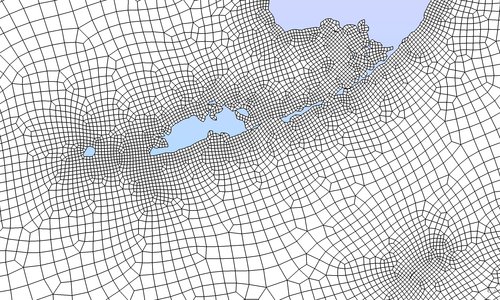}}\\
    \footnotesize{(e) optimization of right-angled triangulation} &
    \footnotesize{(f) final quadrilateral mesh}
  \end{tabular}
  \caption{Step by step generation of a non uniform quadrilateral mesh around the southern tip of Florida peninsula. Generation of points directed by the cross field and optimization of right-angled triangulation maximizes the quality of the output quadrilateral mesh.  \label{fig::quadrangulation}}
  \end{center}
\end{figure}

Figure \ref{fig::quadrangulation} shows the discrete steps of quadrilateral mesh generation workflow. Generation of points guided with a cross field leads to a right-angled triangulation. The optimization process described in section \ref{sec::optimization} is applied to it and further improves the orientation and pairing of right-angled triangles. The final merging to quadrilaterals is done with the blossom-quad method \cite{remacle2012blossom} along with quadrilateral smoothing procedures incorporated in Gmsh \cite{geuzaine2009gmsh}. It must be noted that a small number of remaining triangles along the coastal boundary may remain. This is due to the highly irregular nature of the boundary geometry and can be easily addressed by choosing a mesh size field smaller than the characteristic length of the boundary.

\begin{figure}
  \begin{center}
  \begin{tabular}{cc}
   \setlength{\fboxsep}{0pt} \fbox{\centering \includegraphics[height=7cm]{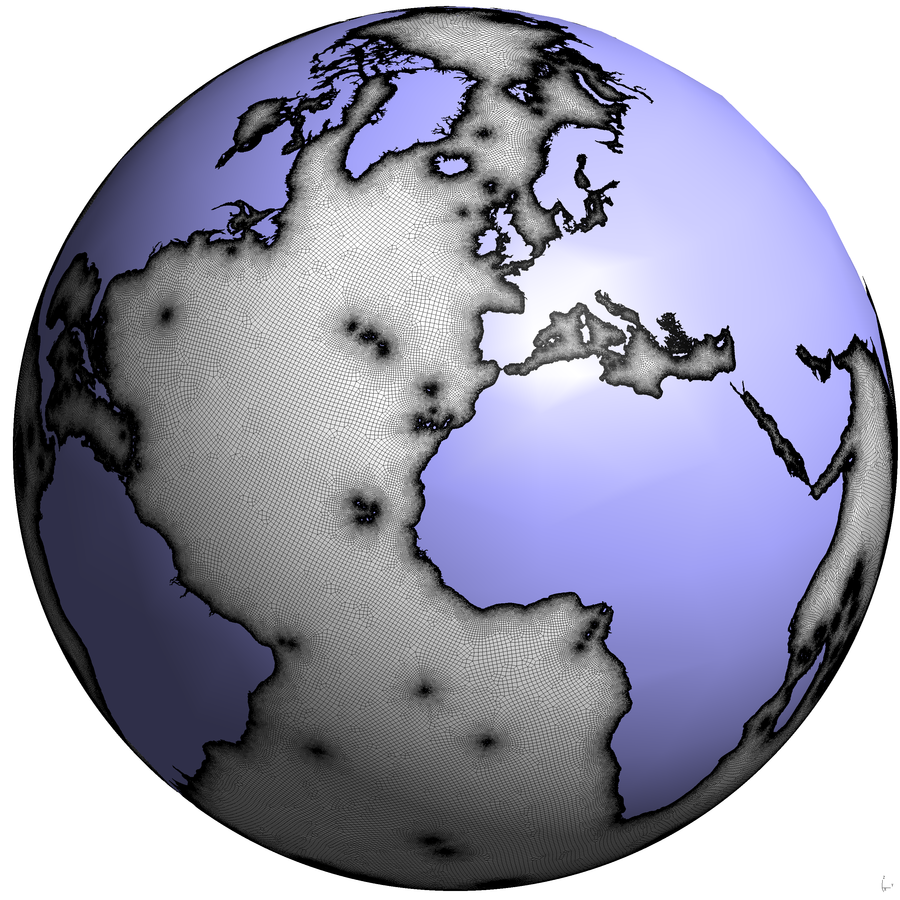}}&
   \setlength{\fboxsep}{0pt} \fbox{\centering \includegraphics[height=7cm]{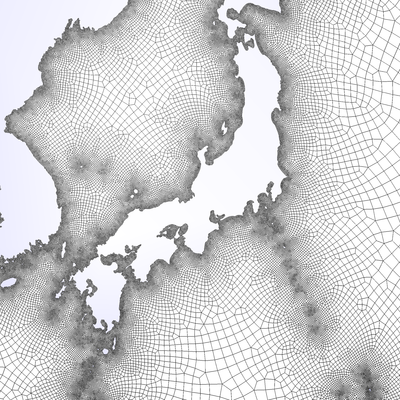}}\\
  \end{tabular}
  \centering
  \setlength{\fboxsep}{0pt} \fbox{\centering \includegraphics[height=8.9cm]{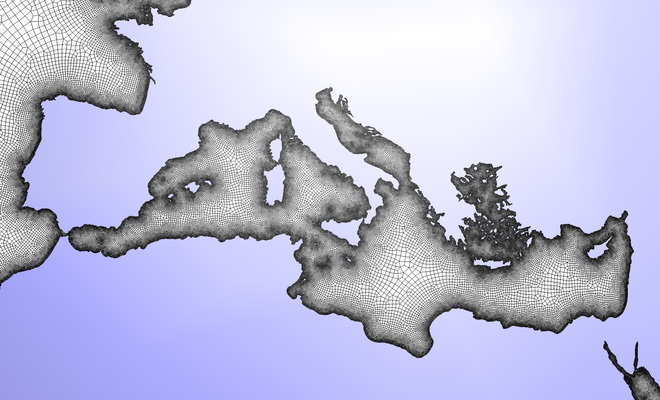}}
  \caption{Quadrilateral mesh of world ocean (top left) with a zoom on the
  Mediterranean sea (bottom)  and on the sea of Japan (top right).\label{fig::world}}
  \end{center}
\end{figure}

Finally, we utilize our algorithm to generate a quadrilateral mesh of the whole world ocean (Figure \ref{fig::world}), with a resolution from $h_{min} = 3$km to $h_{max} = 60$km.  Generation of the final 2,267,738 points takes around 30 seconds and the triangulation of them around 5 seconds. For comparison, the base mesh generated with the procedure of \cite{remacle2016fast} takes around 50 seconds. The average isotropy measure quality \cite{Johnen2016} of the output quadrilateral elements is $0.944$.

\section{Conclusions}

This paper presents a method to generate high quality meshes by utilizing direction fields. By using solely information from an initial triangulation, the algorithm does not require specific knowledge of the geometrical characteristics of the surface. This holds accurately for the case where the mesh size field is the same for the base and the final mesh. 

By utilizing an asterisk field (direction field with 6 turns) we can generate triangulations of high quality in comparison with the initial mesh. Combining this method with an appropriate knowledge of the underlying geometry, it could be used as a tool for re-meshing triangulations to finer resolution and improved quality.

Respectively, by utilizing a cross field (direction field with 4 turns), we can generate right-angled triangulations suitable for recombination to quadrilateral meshes. We have developed an optimization procedure that is applied that is performed on the output right-angled meshes, in order to further maximize the quality of the resultant quadrilaterals.

A multi-threaded strategy for our frontal point generation algorithm is presented. A more detailed analysis of the multi-threaded implementation will follow in the future. The following step of our work is the generation of points in an anisotropic fashion and thus the generation of boundary layer meshes. 

\section*{Acknowledgements}
This research is supported by the European Research Council (project HEXTREME, ERC-2015-AdG-694020).

\bibliographystyle{model1-num-names}
\bibliography{meshing}

\end{document}